\documentclass[10pt,twocolumn,english, aps,superscriptaddress,showpacs,floatfix]{revtex4-1}

\usepackage[T1]{fontenc}
\usepackage[latin9]{inputenc}
\usepackage{color}
\usepackage{babel}
\usepackage{verbatim}
\usepackage{textcomp}
\usepackage{amsmath}
\usepackage{amssymb}
\usepackage{graphicx}
\usepackage[unicode=true,pdfusetitle,
 bookmarks=true,bookmarksnumbered=false,bookmarksopen=false,
 breaklinks=false,pdfborder={0 0 0},pdfborderstyle={},backref=false,colorlinks=true]
 {hyperref}

\makeatletter
\numberwithin{equation}{section}

\usepackage[latin9]{inputenc} 

\usepackage{braket}

\usepackage{amsmath}

\makeatother

\begin{document}
\title{High harmonic generation in crystals using Maximally Localized Wannier
functions}
\author{R. E. F. Silva}
\email{ruiefdasilva@gmail.com}

\affiliation{\emph{Max-Born-Institut, Max Born Strasse 2A, D-12489 Berlin, Germany}}
\affiliation{\emph{Departamento de Física Teórica de la Materia Condensada, Universidad
Autónoma de Madrid, E-28049 Madrid, Spain}}
\author{F. Martín}
\affiliation{\emph{Departamento de Qu\'{\i}mica, Universidad Autónoma de Madrid,
28049 Madrid, Spain}}
\affiliation{\emph{Instituto Madrileño de Estudios Avanzados en Nanociencia, 28049
Madrid, Spain}}
\affiliation{\emph{Condensed Matter Physics Center (IFIMAC), Universidad Autónoma
de Madrid, E-28049 Madrid, Spain}}
\author{M. Ivanov}
\email{mikhail.ivanov@mbi-berlin.de}

\affiliation{\emph{Max-Born-Institut, Max Born Strasse 2A, D-12489 Berlin, Germany}}
\affiliation{\emph{Blackett Laboratory, Imperial College London, South Kensington
Campus, SW7 2AZ London, United Kingdom}}
\affiliation{\emph{Department of Physics, Humboldt University, Newtonstrasse 15,
12489 Berlin, Germany}}
\begin{abstract}
In this work, the nonlinear optical response, and in particular, the
high harmonic generation of semiconductors is addressed by using the
Wannier gauge. One of the main problems in the time evolution of the
Semiconductor Bloch equations resides in the fact that the dipole
couplings between different bands can diverge and have a random phase
along the reciprocal space and this leads to numerical instability.
To address this problem, we propose the use of the Maximally Localized
Wannier functions that provide a framework to map \emph{ab-initio}
calculations to an effective tight-binding Hamiltonian with great
accuracy. We show that working in the Wannier gauge, the basis set
in which the Bloch functions are constructed directly from the Wannier
functions, the dipole couplings become smooth along the reciprocal
space thus avoiding the problem of random phases. High harmonic generation
spectrum is computed for a 2D monolayer of hBN as a numerical demonstration.%
\end{abstract}
\maketitle

\section{Introduction}

Since the discovery of high harmonic generation (HHG) in solids \citep{ghimire2011observation},
several works demonstrate the potential of analyzing the light emitted
by bulk electrons in solids when exposed to intense laser fields.
Recent results include the observation of dynamical Bloch oscillations
\citep{schubert2014sub,luu2015extreme}, band structure tomography
\citep{vampa2015all,tancogne2017ellipticity},  resolving electron-hole
dynamics \citep{bauer2018high,mcdonald2015interband}, including dynamics
in strongly correlated systems and phase transitions in the Mott insulator
\citep{silva2018high}, the Peierls phase transition \citep{bauer2018high},
or the imprint of the Berry phase on high harmonic spectrum \citep{luu2018measurement,liu2017high}. 

The first theoretical works trying to describe HHG in solids are previous
to the experimental realization of the process \citep{plaja1992high,golde2008high}.
But the interest in the theoretical understanding had grown in the
last years \citep{hawkins2013role,Vampa2014,hawkins2015effect,Korbman2013,wismer2016strong,floss2018ab,osika2017wannier,tancogne2017ellipticity}.
Several theoretical approaches were used in this context: solution
of the time dependent Schrödinger equation (TDSE) \citep{Korbman2013},
time-dependent density functional theory (TDDFT) \citep{tancogne2017ellipticity}
and semiconductor Bloch equations (SBE) \citep{kira_koch_book,golde2008high}.

Despite the variety of theoretical approaches used up to now, the
calculation of nonlinear optical properties of crystals presents several
difficulties. In the dipole approximation, two different gauges for
the electromagnetic field are usually used: length gauge (LG) and
velocity gauge (VG). In the VG, the advantage with respect to the
LG is that the dynamical equations become decoupled in the Brillouin
zone but a large number of bands need to be included in the calculation
to obtain the converged result \citep{aversa1995nonlinear,ventura2017gauge}.
On the other hand, the LG gives the converged result for low frequencies
with a modest number of bands included in the calculation. In the
LG, the problem arises in the representation of the position operator,
that now involves a derivative in the Brillouin zone \citep{blount1962formalisms}.
Evaluating this derivative is a numerical challenge since the Bloch
eigenstates of the crystalline Hamiltonian are defined up to a random
phase in the Brillouin zone. In this work, we propose a way to circumvent
this problem by using the Maximally Localized Wannier functions (MLWF)
that cure this gauge freedom \citep{marzari1997maximally,marzari2012maximally}.
The formalism developed in this work can be applied to analytical
tight-binding models but also for DFT calculations that can be mapped
to a tight-binding Hamiltonian using the MLWF procedure \citep{mostofi2014updated,marzari1997maximally,marzari2012maximally}.

\section{Theory}

In this section, we give a brief overview of the theory of nonlinear
optical response in semiconductors. It is worth to point out the reader
to the work of Blount \citep{blount1962formalisms} and to the work
of Ventura \emph{et al. }\citep{ventura2017gauge,ventura2016masterthesis}
where detailed discussion is given on the difficulties that the position
operator presents in a crystal. We derive the dynamical equations
for the calculation of the bulk current of a semiconductor when in
presence of an intense laser pulse. We will work in the length gauge
(LG) under the dipole approximation and neglecting electron-electron
and electron-phonon interactions.

The novelty of this work relies on the numerical solution of the dynamical
equations in a diabatic basis instead of solving them in an adiabatic
basis. As it is well know in molecular physics, the study of molecular
dynamics is sometimes cumbersome if one chooses to work with an adiabatic
basis, especially when one needs to deal with avoided crossings or
conical intersections. It is much more appropriate to study molecular
dynamics in a diabatic basis, in which non-adiabatic couplings are
absent, or as small as possible \citep{tannor2006book}. Two major
problems arise when calculating the non-adiabatic couplings in the
adiabatic basis: the gauge freedom for choosing the eigenstates of
the Hamiltonian up to a random phase and the divergences of the non-adiabatic
couplings in the vicinity of the avoided crossings or conical intersections.

In a complete analogy with molecular physics, this work proposes a
diabatic basis in which the numerical solution of SBE can be achieved
easily, the Wannier gauge \citep{marzari2012maximally}. It must be
stressed out that when one is referring to the LG versus the VG, we
are dealing with the intrinsic gauge freedom to describe the electromagnetic
field. On the other hand, when we are referring to the Hamiltonian
gauge versus the Wannier gauge, we are dealing with two different
basis set in which we can solve our quantum mechanical problem. The
Hamiltonian gauge is a choice of a basis set in which we expand our
wavefunction (or density matrix) in Bloch eigenstates of the crystalline
Hamiltonian. On the other hand, the Wannier gauge is the gauge in
which calculations are done using an expansion on Bloch states that
are constructed from localized Wannier orbitals. In analogy with molecular
physics, the Hamiltonian gauge is what we would think as an adiabatic
basis and the Wannier gauge is what we would think as a diabatic basis.

\subsection{Single-Particle Theory}

Within the dipole approximation and in the LG, the time-dependent
Hamiltonian describing the interaction of an electron in a periodic
potential with a laser field is 
\begin{align}
\hat{H}\left(t\right) & =\hat{H}_{0}+\left|e\right|\boldsymbol{E}\left(t\right).\hat{\boldsymbol{r}}\\
\hat{H}_{0} & =\frac{\boldsymbol{p}^{2}}{2m_{e}}+U\left(\boldsymbol{r}\right)\\
U\left(\boldsymbol{r}\right) & =U\left(\boldsymbol{r}+\boldsymbol{R}\right)
\end{align}
where $\boldsymbol{E}\left(t\right)$ is the electric field, $\left|e\right|$
is the elementary charge, $m_{e}$ is the electron mass and $\boldsymbol{R}$
is a Bravais lattice vector. Let us consider a set of $M$ localized
Wannier orbitals in each cell $w_{m}\left(\boldsymbol{r}-\boldsymbol{R}\right)=\Braket{\boldsymbol{r}|\boldsymbol{R}m}$.
We assume that these Wannier orbitals form an orthonormal basis, i.e.
$\Braket{\boldsymbol{R}'n|\boldsymbol{R}m}=\delta_{n,m}\delta_{\boldsymbol{R'},\boldsymbol{R}}$.
Assuming that our system contains $N_{c}$ unitary cells with periodic
boundary conditions, we can define Bloch like functions from these
Wannier orbitals

\begin{align}
\Ket{\psi_{m\boldsymbol{k}}^{\left(W\right)}} & =\frac{1}{\sqrt{N_{c}}}\sum_{\boldsymbol{R}}e^{i\boldsymbol{k}.\boldsymbol{R}}\Ket{\boldsymbol{R}m}\\
\Braket{\boldsymbol{r}|\psi_{m\boldsymbol{k}}^{\left(W\right)}} & =\frac{1}{\sqrt{N_{c}V_{c}}}e^{i\boldsymbol{k}.\boldsymbol{r}}u_{m\boldsymbol{k}}^{\left(W\right)}\left(\boldsymbol{r}\right)
\end{align}
where $V_{c}$ is the volume of a unit cell and $u_{m\boldsymbol{k}}^{\left(W\right)}\left(\boldsymbol{r}\right)$
is the periodic part of the Bloch function.

We are interested in the calculation of nonlinear optical response
within the dipole approximation and for a complete description of
the light-matter interaction it is only necessary to have the knowledge
of
\begin{align}
\Braket{\boldsymbol{0}n|\hat{H}_{0}|\boldsymbol{R}m}\\
\Braket{\boldsymbol{0}n|\hat{\boldsymbol{r}}|\boldsymbol{R}m}
\end{align}
where $\hat{H}_{0}$ is the unperturbed Hamiltonian and $\hat{\boldsymbol{r}}$
is the position operator. These matrix elements can either be calculated
from an electronic structure calculation followed by a wannierization
procedure \citep{marzari2012maximally} or can be set explicitly by
analytical tight-binding models. We can define, as in \citep{wang2006ab},
the Hamiltonian and Berry connection matrices in the Wannier gauge
\begin{align}
H_{nm}^{(W)}\left(\boldsymbol{k}\right) & \equiv\sum_{\boldsymbol{R}}e^{i\boldsymbol{k}.\boldsymbol{R}}\left\langle \boldsymbol{0}n\left|\hat{H}_{0}\right|\boldsymbol{R}m\right\rangle \label{eq:Ham_Wannier}\\
\boldsymbol{A}_{nm}^{(W)}\left(\boldsymbol{k}\right) & \equiv\sum_{\boldsymbol{R}}e^{i\boldsymbol{k}.\boldsymbol{R}}\left\langle \boldsymbol{0}n\left|\hat{\boldsymbol{r}}\right|\boldsymbol{R}m\right\rangle .\label{eq:Berry_connection_Wannier}
\end{align}
where the sum in $\boldsymbol{R}$ runs over all lattice vectors in
our system and the superscript $\left(W\right)$ refers to the Wannier
gauge (as in \citep{wang2006ab}). As it will be seen later, these
matrices enter in the dynamical equations of our system. Since we
are working under the assumption that the Wannier orbitals are localized
functions, we expect that the terms, $\left\langle \boldsymbol{0}n\left|\hat{H}_{0}\right|\boldsymbol{R}m\right\rangle $
and $\left\langle \boldsymbol{0}n\left|\hat{\boldsymbol{r}}\right|\boldsymbol{R}m\right\rangle $,
in the sum will decay exponentially with $\boldsymbol{R}$ and this
fact will lead to smooth matrices, $H^{(W)}\left(\boldsymbol{k}\right)$
and $\boldsymbol{A}^{(W)}\left(\boldsymbol{k}\right)$.

We can write the single particle unperturbed Hamiltonian as 
\begin{align}
\hat{H}_{0} & =\sum_{\boldsymbol{k}\in FBZ}\sum_{n,m}H_{nm}^{(W)}\left(\boldsymbol{k}\right)\Ket{\psi_{n\boldsymbol{k}}^{\left(W\right)}}\Bra{\psi_{m\boldsymbol{k}}^{\left(W\right)}}
\end{align}
where the sum over $\boldsymbol{k}$ runs over all allowed crystal
momentum inside the first Brillouin zone (FBZ). We can work in the
basis where $\hat{H}_{0}$ is diagonal and for that we only need to
diagonalize the $H_{nm}^{(W)}\left(\boldsymbol{k}\right)$ matrix.
We will refer to this basis as the Hamiltonian gauge using the superscript
$\left(H\right)$. Whenever none of the superscripts are used it means
that the result is valid for both, $\left(W\right)$ and $\left(H\right)$,
gauges. Both basis are related to each other by a unitary transformation
\begin{equation}
\ket{\psi_{n\boldsymbol{k}}^{\left(H\right)}}=\sum_{m}U_{mn}\left(\boldsymbol{k}\right)\ket{\psi_{m\boldsymbol{k}}^{\left(W\right)}}
\end{equation}
where $U\text{\ensuremath{\left(\boldsymbol{k}\right)}}$ is a unitary
matrix that diagonalizes $H^{(W)}\left(\boldsymbol{k}\right)$,
\begin{equation}
U^{\dagger}\text{\ensuremath{\left(\boldsymbol{k}\right)}}H^{(W)}\left(\boldsymbol{k}\right)U\text{\ensuremath{\left(\boldsymbol{k}\right)}}=H^{(H)}\left(\boldsymbol{k}\right).
\end{equation}
The relationship between the $\boldsymbol{A}\left(\boldsymbol{k}\right)$
matrix in both gauges is \citep{wang2006ab}
\begin{equation}
\boldsymbol{A}^{\left(H\right)}\left(\boldsymbol{k}\right)=U^{\dagger}\text{\ensuremath{\left(\boldsymbol{k}\right)}}\boldsymbol{A}^{\left(W\right)}\left(\boldsymbol{k}\right)U\text{\ensuremath{\left(\boldsymbol{k}\right)}}+iU^{\dagger}\text{\ensuremath{\left(\boldsymbol{k}\right)}}\frac{\partial}{\partial\boldsymbol{k}}U\left(\boldsymbol{k}\right).
\end{equation}
It must be noted that the matrix $U\left(\boldsymbol{k}\right)$ is
not unique. This fact lies on the gauge freedom that we have to choose
a global phase for the eigenstates, $\Ket{\psi_{n\boldsymbol{k}}^{\left(H\right)}}\rightarrow e^{i\theta_{n}\left(\boldsymbol{k}\right)}\Ket{\psi_{n\boldsymbol{k}}^{\left(H\right)}}$.
Since the last term, $iU^{\dagger}\text{\ensuremath{\left(\boldsymbol{k}\right)}}\frac{\partial}{\partial\boldsymbol{k}}U\left(\boldsymbol{k}\right)$,
is not invariant under this phase transformation, we have an intrinsic
gauge freedom in the calculation of $\boldsymbol{A}^{\left(H\right)}\left(\boldsymbol{k}\right)$.
It is precisely this gauge freedom that we try to avoid using the
Wannier gauge.

The matrix elements of the position operator in a finite volume system
with periodic boundary conditions are ill-defined, but can be computed
in the thermodynamic limit. The representation of the position operator
in a Bloch basis is \citep{blount1962formalisms,ventura2017gauge}
\begin{equation}
\hat{\boldsymbol{r}}=i\frac{\partial}{\partial\boldsymbol{k}}+\hat{\boldsymbol{A}}
\end{equation}
where $\hat{\boldsymbol{A}}$ is the Berry connection that is defined
as
\begin{equation}
\boldsymbol{A}_{nm}\left(\boldsymbol{k}\right)=i\int_{V_{C}}d^{D}\boldsymbol{r}\left(u_{n\boldsymbol{k}}\left(\boldsymbol{r}\right)\right)^{*}\frac{\partial}{\partial\boldsymbol{k}}u_{m\boldsymbol{k}}\left(\boldsymbol{r}\right).
\end{equation}
In the Wannier gauge the Berry connection will be the one defined
in Eq. (\ref{eq:Berry_connection_Wannier}).

The matrix representation of the current operator in a Bloch basis
is defined as 
\begin{align}
\hat{\boldsymbol{J}} & =\frac{-\left|e\right|}{i\hbar}\left[\hat{\boldsymbol{r}},\hat{H}\left(t\right)\right]\\
\left(\hat{\boldsymbol{J}}\right)_{\boldsymbol{k},nm} & =\frac{-\left|e\right|}{\hbar}\left(\frac{\partial}{\partial\boldsymbol{k}}H_{nm}\left(\boldsymbol{k}\right)-i\left[\boldsymbol{A}\left(\boldsymbol{k}\right),H\left(\boldsymbol{k}\right)\right]_{nm}\right).
\end{align}
In the Hamiltonian gauge, we can identify the two terms in the expression
for the current as the \emph{intraband }and \emph{interband }current.
\begin{align}
\left(\hat{\boldsymbol{J}}_{intra}\right)_{\boldsymbol{k},nm} & =\frac{-\left|e\right|}{\hbar}\left(\frac{\partial}{\partial\boldsymbol{k}}H_{nm}^{\left(H\right)}\left(\boldsymbol{k}\right)\right)\\
\left(\hat{\boldsymbol{J}}_{inter}\right)_{\boldsymbol{k},nm} & =\frac{i\left|e\right|}{\hbar}\left(\left[\boldsymbol{A}^{\left(H\right)}\left(\boldsymbol{k}\right),H^{\left(H\right)}\left(\boldsymbol{k}\right)\right]_{nm}\right).
\end{align}

\subsection{Many-body theory}

Having all the relevant operators (the current and the time-dependent
Hamiltonian operator) defined in first quantization, we proceed to
define all the relevant operators in a second quantization formalism.
The time-dependent many-body Hamiltonian is just
\begin{align}
\hat{\mathcal{H}}\left(t\right) & =\sum_{\boldsymbol{k}\in FBZ}\sum_{n,m}c_{n\boldsymbol{k}}^{\dagger}H_{nm}\left(\boldsymbol{k}\right)c_{m\boldsymbol{k}}\nonumber \\
 & +\sum_{\boldsymbol{k}\in FBZ}\sum_{n,m}c_{n\boldsymbol{k}}^{\dagger}\left|e\right|\boldsymbol{E}\left(t\right).\left[i\delta_{nm}\frac{\partial}{\partial\boldsymbol{k}}+\boldsymbol{A}_{nm}\left(\boldsymbol{k}\right)\right]c_{m\boldsymbol{k}}
\end{align}
where $c_{n\boldsymbol{k}}^{\dagger}$ ($c_{n\boldsymbol{k}}$) is
the fermionic creation (annihilation) operator of a Bloch state. The
many-body current operator can also be expressed as 
\begin{align}
\hat{\boldsymbol{\mathcal{J}}} & =\sum_{\boldsymbol{k}\in FBZ}\sum_{n,m}\left(\hat{\boldsymbol{J}}\right)_{\boldsymbol{k},nm}c_{n\boldsymbol{k}}^{\dagger}c_{m\boldsymbol{k}}\\
 & =\sum_{\boldsymbol{k}\in FBZ}\sum_{n,m}\frac{-\left|e\right|}{\hbar}\left(\frac{\partial}{\partial\boldsymbol{k}}H_{nm}\left(\boldsymbol{k}\right)\right)c_{n\boldsymbol{k}}^{\dagger}c_{m\boldsymbol{k}}\nonumber \\
 & +\sum_{\boldsymbol{k}\in FBZ}\sum_{n,m}\frac{i\left|e\right|}{\hbar}\left[\boldsymbol{A}\left(\boldsymbol{k}\right),H\left(\boldsymbol{k}\right)\right]_{nm}c_{n\boldsymbol{k}}^{\dagger}c_{m\boldsymbol{k}}
\end{align}
The observables that we are interested in is the mean value of the
current operator, $\hat{\left\langle \boldsymbol{\mathcal{J}}\right\rangle }$,
and for that we only need to know the mean values of $c_{n\boldsymbol{k}}^{\dagger}c_{m\boldsymbol{k}}$.
Let us define the reduced density matrix (RDM) as
\begin{equation}
\rho_{nm}\left(\boldsymbol{k},t\right)=\left\langle c_{m\boldsymbol{k}}^{\dagger}c_{n\boldsymbol{k}}\right\rangle .
\end{equation}
Note that the band indexes are switched. The equations of motion for
the RDM have the following form \citep{ventura2017gauge}
\begin{align}
i\hbar\frac{\partial}{\partial t}\rho_{nm}\left(\boldsymbol{k},t\right) & =\left\langle \left[c_{m\boldsymbol{k}}^{\dagger}c_{n\boldsymbol{k}},\hat{\mathcal{H}}\left(t\right)\right]\right\rangle \nonumber \\
 & =\left[H\left(\boldsymbol{k}\right),\rho\left(\boldsymbol{k},t\right)\right]_{nm}\nonumber \\
 & +i\left|e\right|\boldsymbol{E}\left(t\right).\frac{\partial}{\partial\boldsymbol{k}}\rho_{nm}\left(\boldsymbol{k},t\right)\nonumber \\
 & +\left|e\right|\boldsymbol{E}\left(t\right).\left[\boldsymbol{A}\left(\boldsymbol{k}\right),\rho\left(\boldsymbol{k},t\right)\right]_{nm}\label{eq:RDM_eq_motion}
\end{align}
With these equations we have the complete set of dynamical equations
for our problem. It is precisely when we look at the structure of
the equations of motion for the RDM that we can appreciate the advantage
of using the Wannier gauge. Since the Wannier gauge provides smooth
$\boldsymbol{A}\left(\boldsymbol{k}\right)$ and $H\left(\boldsymbol{k}\right)$,
the numerical propagation of the equations of motion in a discretized
$\boldsymbol{k}$-grid is easier compared to the Hamiltonian gauge
case.

\subsection{Initial Conditions}

We can assume that the electrons in a solid are in thermal equilibrium
before the application of a coherent perturbation. Thus our initial
state is a mixed state without coherence between eigenstates with
a Fermi-Dirac distribution. We can construct the unperturbed RDM in
the $\left(H\right)$ gauge
\begin{equation}
\rho_{nm}^{\left(H\right)}\left(\boldsymbol{k},t_{0}\right)=\delta_{nm}s_{n}p_{n\boldsymbol{k}}
\end{equation}
where $s_{n}=1,2$ takes into account spin degeneracy, $p_{n\boldsymbol{k}}=1/\left(e^{\left(\left(\varepsilon_{n}\left(\boldsymbol{k}\right)-\mu\right)/k_{B}T\right)}\right)$
and $\mu$ is the chemical potential. The relationship between the
two RDM in both gauges is given by 
\begin{align}
\rho_{nm}^{\left(H\right)}\left(\boldsymbol{k},t\right) & =\sum_{ab}U_{nb}^{\dagger}\left(\boldsymbol{k}\right)\rho_{ba}^{\left(W\right)}\left(\boldsymbol{k},t\right)U_{am}\left(\boldsymbol{k}\right)\\
\rho_{nm}^{\left(W\right)}\left(\boldsymbol{k},t\right) & =\sum_{ab}U_{nb}\left(\boldsymbol{k}\right)\rho_{ba}^{\left(H\right)}\left(\boldsymbol{k},t\right)U_{am}^{\dagger}\left(\boldsymbol{k}\right).
\end{align}
Since the initial state at $t_{0}$ does not have any coherence (i.e.
when $n\neq m$, $\left\langle c_{m\boldsymbol{k}}^{\left(H\right)\dagger}c_{n\boldsymbol{k}}^{\left(H\right)}\right\rangle =0$)
the gauge freedom in the choice of $U$ will not change $\rho_{nm}^{\left(W\right)}\left(\boldsymbol{k},t_{0}\right)$.
This makes our initial RDM in the Wannier gauge robust under phase
transformations.

\subsection{Dephasing}

A single electron picture is not enough to describe nonlinear optical
response in solids and for a proper modeling of the HHG process and
decoherence due to electron-electron or electron-phonon scattering
must be taken into account. In this work, we will just introduce decoherence
in the same way it was introduced by Vampa \emph{et al.} \citep{Vampa2014}
in the Hamiltonian gauge. There is an ongoing debate about the origin
of the very short dephasing times needed to obtain agreement with
the experimental results ($T_{2}\approx2$fs) \citep{Vampa2014}.
Recently, the importance of propagation effects in solids was pointed
out in\emph{ }\citep{floss2018ab}. The dynamical equation for the
RDM in the Hamiltonian gauge including dephasing will now take the
form
\begin{align}
i\hbar\frac{\partial}{\partial t}\rho_{nm}^{\left(H\right)}\left(\boldsymbol{k},t\right) & =\left[H^{\left(H\right)}\left(\boldsymbol{k}\right),\rho^{\left(H\right)}\left(\boldsymbol{k},t\right)\right]_{nm}\nonumber \\
 & +i\left|e\right|\boldsymbol{E}\left(t\right).\frac{\partial}{\partial\boldsymbol{k}}\rho_{nm}^{\left(H\right)}\left(\boldsymbol{k},t\right)\nonumber \\
 & +\left|e\right|\boldsymbol{E}\left(t\right).\left[\boldsymbol{A}^{\left(H\right)}\left(\boldsymbol{k}\right),\rho^{\left(H\right)}\left(\boldsymbol{k},t\right)\right]_{nm}\nonumber \\
 & -i\hbar\frac{\left(1-\delta_{nm}\right)}{T_{2}}\rho_{nm}^{\left(H\right)}\left(\boldsymbol{k},t\right).
\end{align}
The dephasing term in the equation of motion for the RDM is better
to be handled in the Hamiltonian gauge. In practice, to solve this
set of equations numerically we propagate the coherent part of the
equation in the Wannier gauge and the dephasing term is introduced
in the Hamiltonian gauge.

\section{Numerical Results}

\begin{figure}
\begin{centering}
\includegraphics[width=1\columnwidth]{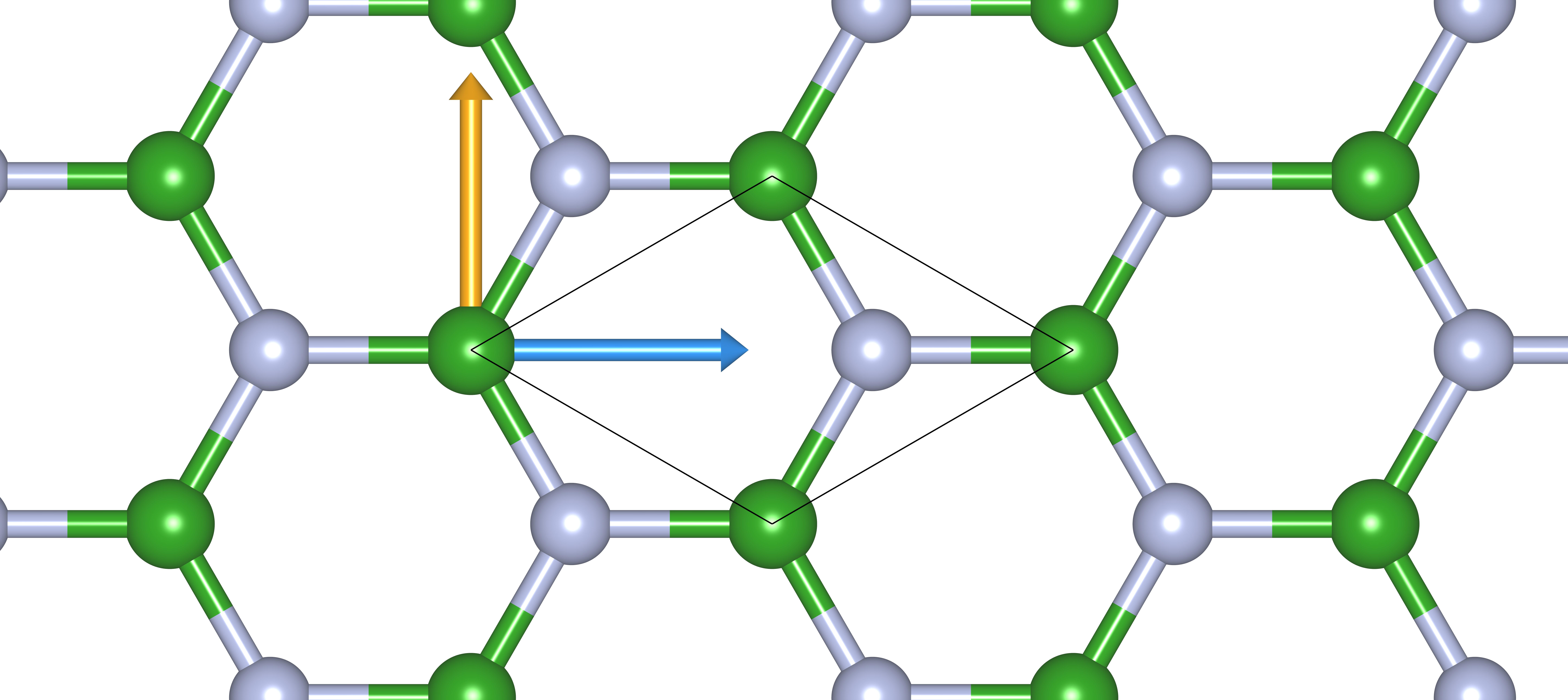}
\par\end{centering}
\raggedright{}\caption{\label{fig:1}Single layer of hexagonal boron nitride. The boron (nitrogen)
atoms are in green (grey). The blue (yellow) arrow represents the
laser polarization along $\Gamma-\mathrm{M}$ ($\Gamma-\mathrm{K}$).}
\end{figure}

In this section, we will show the results of applying the formalism
developed to the case of a single monolayer of hexagonal boron nitride
(hBN). In Fig. \ref{fig:1}, we show the hexagonal structure of the
hBN and the two laser polarizations that will be used in this work,
along $\Gamma-\mathrm{M}$ and $\Gamma-\mathrm{K}$ directions. The
distance between neighbor atoms is 1.446 $\text{Å}$. We will use
two different set of parameters for the Hamiltonian and dipole couplings.

\begin{figure}
\begin{centering}
\includegraphics[width=1\columnwidth]{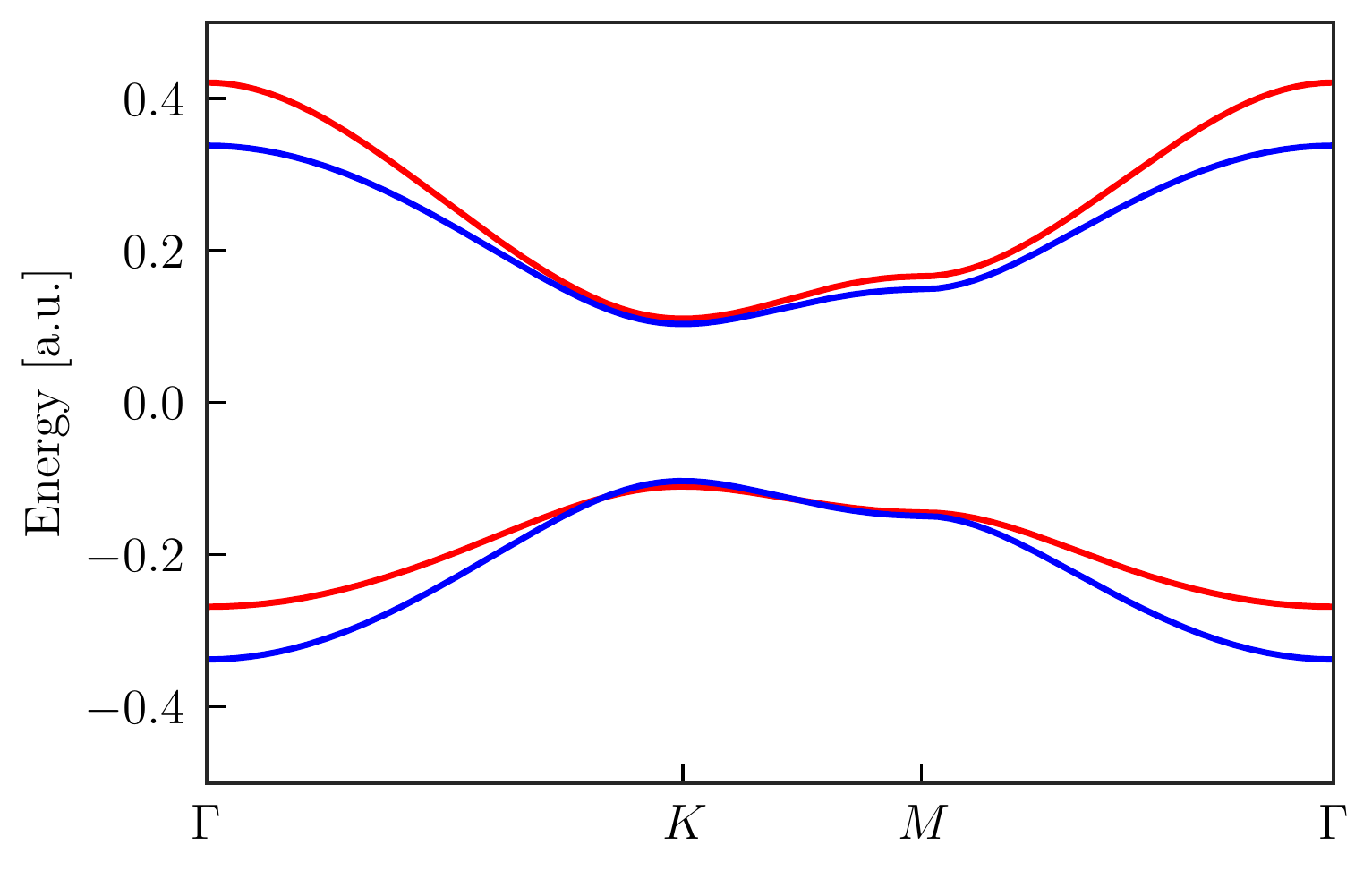}
\par\end{centering}
\raggedright{}\caption{\label{fig:2}Electronic band structure for the effective model (blue
line) and for the \emph{ab-initio} model (red line).}
\end{figure}

\begin{figure}
\begin{centering}
\includegraphics[width=0.33333\textwidth]{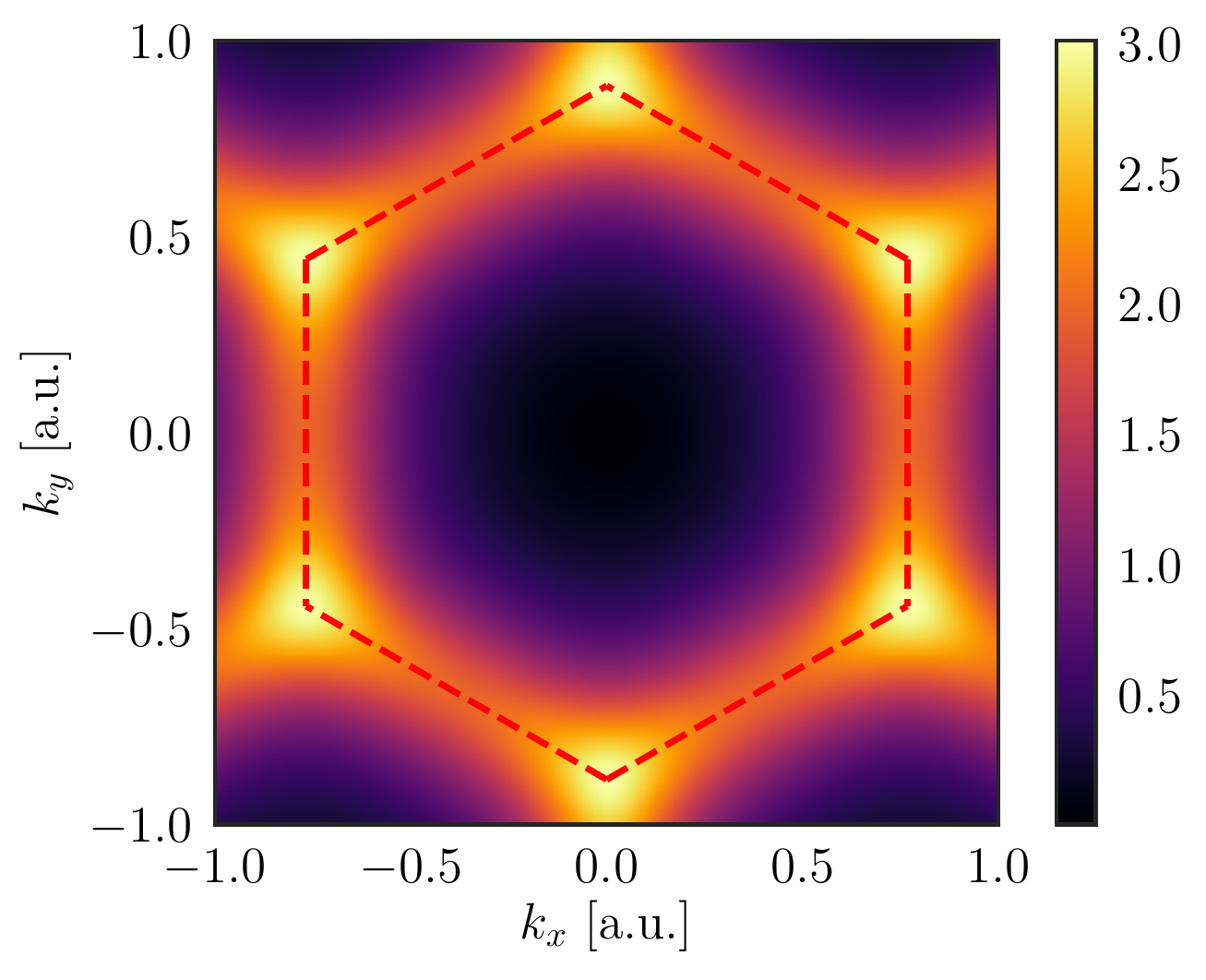}
\par\end{centering}
\raggedright{}\caption{\label{fig:3}Berry connection between the valence band and the conduction
band, $\left|\boldsymbol{A}_{vc}^{\left(H\right)}\left(\boldsymbol{k}\right)\right|=\sqrt{\left|A_{x,vc}^{\left(H\right)}\left(\boldsymbol{k}\right)\right|^{2}+\left|A_{y,vc}^{\left(H\right)}\left(\boldsymbol{k}\right)\right|^{2}}$,
for the \emph{ab-initio} model.}
\end{figure}

\begin{figure}
\begin{centering}
\includegraphics[width=0.33333\textwidth]{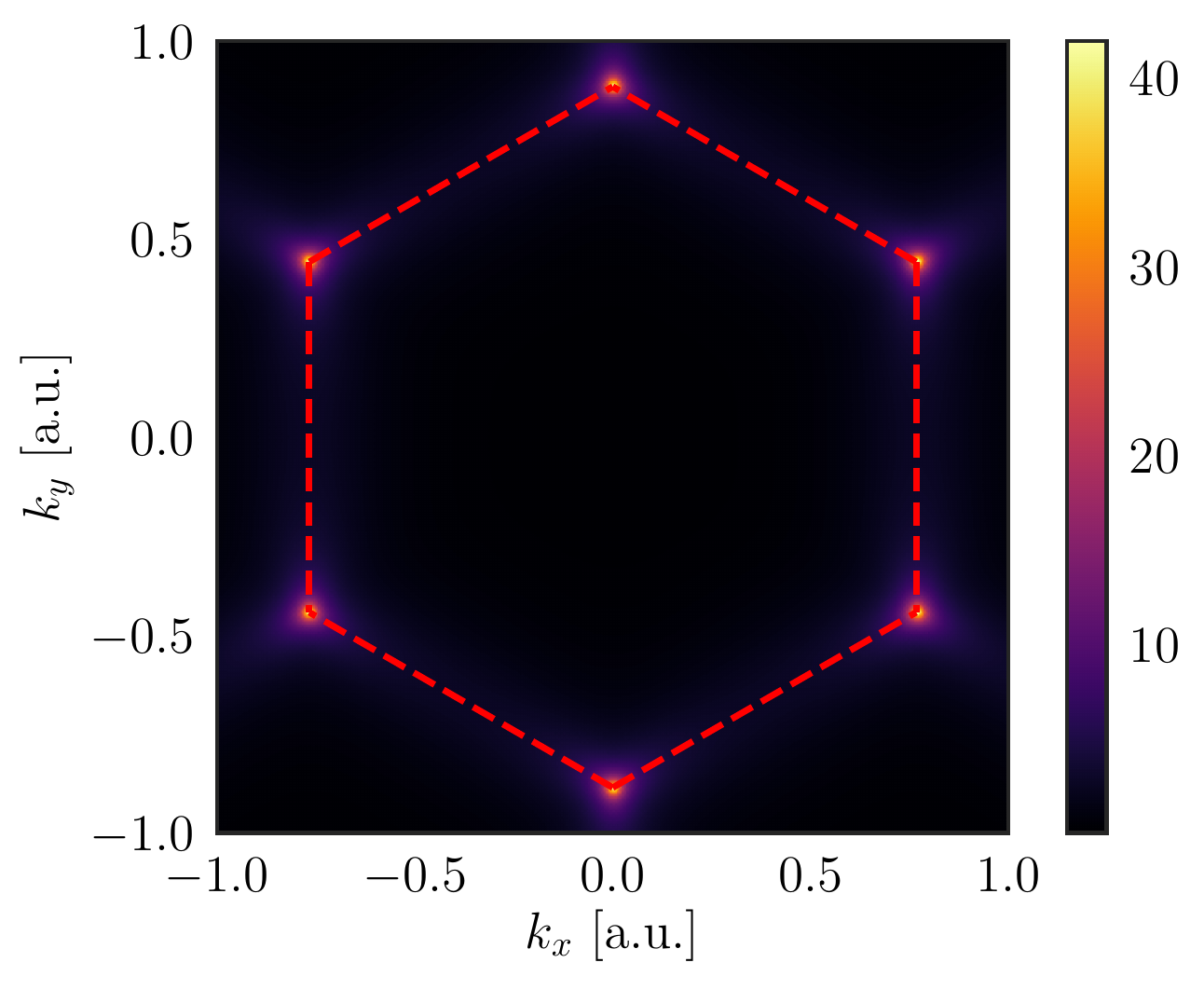}
\par\end{centering}
\raggedright{}\caption{\label{fig:4}Same as in Fig. \ref{fig:3} but for $\varepsilon_{B}=-\varepsilon_{N}=0.2$
eV.}
\end{figure}

The first one is the widely used simple tight-binding model in which
only $p_{z}$ orbitals are considered \citep{neto2009electronic}.
We set the hopping constant, $t_{0}$, to be 2.92 eV and the on-site
energy of the two different atoms to be $\varepsilon_{B}=-\varepsilon_{N}=2.81$
eV. For the position operator we assume that it is diagonal in the
basis of the localized Wannier orbitals, i.e. 
\begin{equation}
\left\langle \boldsymbol{0}n\left|\hat{\boldsymbol{r}}\right|\boldsymbol{R}m\right\rangle =\delta_{nm}\delta_{\boldsymbol{0}\boldsymbol{R}}\boldsymbol{\tau}_{n},\label{eq:digonal_r}
\end{equation}
where $\boldsymbol{\tau}_{n}$ is the center of the $n$ Wannier function.
We will refer to this parametrization as effective model. The second
parametrization is done by performing an \emph{ab-initio} calculation
using the HSE06 functional and a 10$\times$10 Monkhorst-Pack grid
using the QuantumEspresso code \citep{giannozzi2009quantum}. We perform
a projection on the $p_{z}$ orbitals and a wannierization procedure
to obtain the Hamiltonian and dipole couplings using the Wannier90
software \citep{mostofi2008wannier90}. We will refer to this parametrization
as \emph{ab-initio} model. In Fig. \ref{fig:2} we show the electronic
band structure for both models.

In order to emphasize the necessity of employing the Wannier gauge
for the numerical computation of the nonlinear optical properties,
we show in Fig. \ref{fig:3} the Berry connection between the valence
and the conduction band. It is clear that the Berry connection is
centered around the $\mathrm{K}$ and $\mathrm{K}'$ points and when
the on-site energy of the atoms goes to smaller values, see Fig. \ref{fig:4},
the Berry connection gets more peaked. To properly describe these
abrupt changes in the BZ using the Hamiltonian gauge requires a very
fine grid in \emph{k}-space. On the other hand, in the Wannier gauge
the Bloch functions are defined to be smooth and continuous in the
reciprocal space. Another advantage of using the Wannier gauge is
that in the case of using an \emph{ab-initio} calculation we can circumvent
the problem of random phases \citep{yu2019high}.

The laser pulse has a peak field of 40 MV/cm, a wavelength of 3 $\mu$m
and 34.2 fs of FWHM in intensity with a $\cos^{2}$ envelope. We solve
the equation of motion for the RDM using a 300$\times$300 Monkhorst-Pack
grid and propagating with a fourth order Runge-Kutta propagator with
a timestep of 2.5 as. For all the calculations, we have used a dephasing
time, $T_{2}$, of 2 fs. Two different polarizations for the electric
field of the laser pulse will be used, along the $\Gamma-\mathrm{M}$
(0º) and along the $\Gamma-\mathrm{K}$ (90º) direction. The HHG spectrum
will be evaluated along the parallel ($\parallel$) and perpendicular
($\perp$) direction of the laser pulse.

\begin{figure*}[t]
\begin{centering}
\includegraphics[width=0.333333\textwidth]{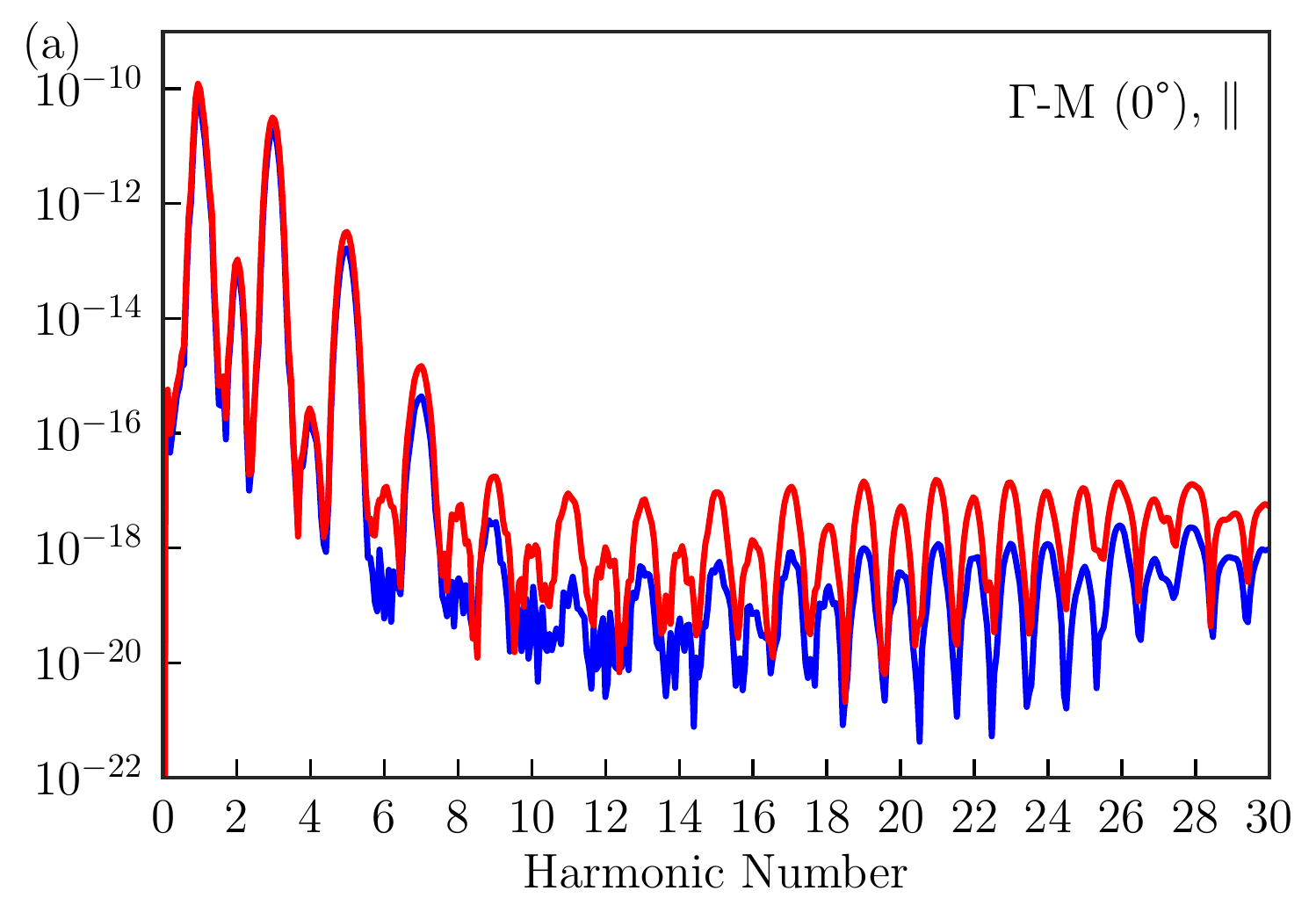}\includegraphics[width=0.333333\textwidth]{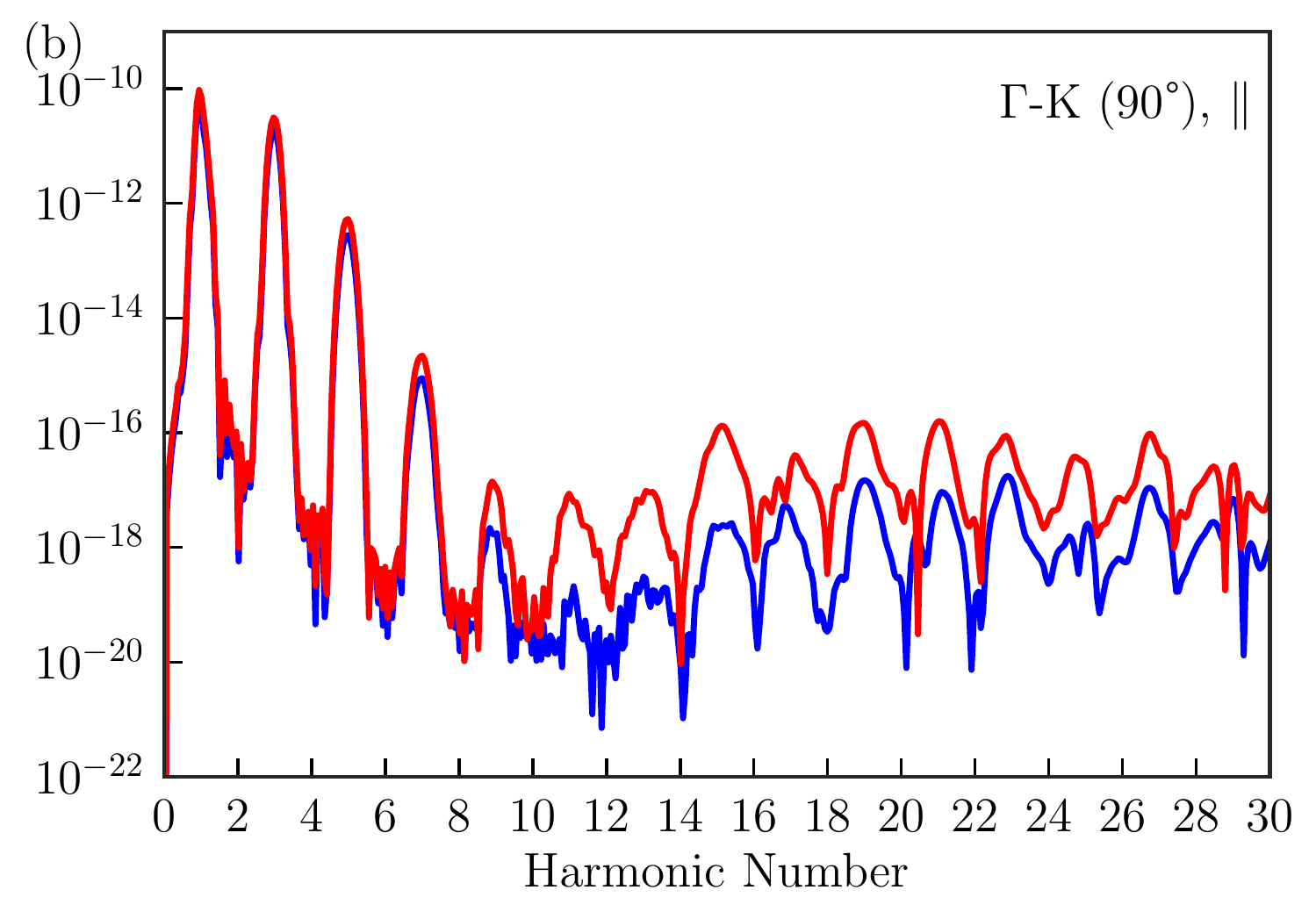}\includegraphics[width=0.333333\textwidth]{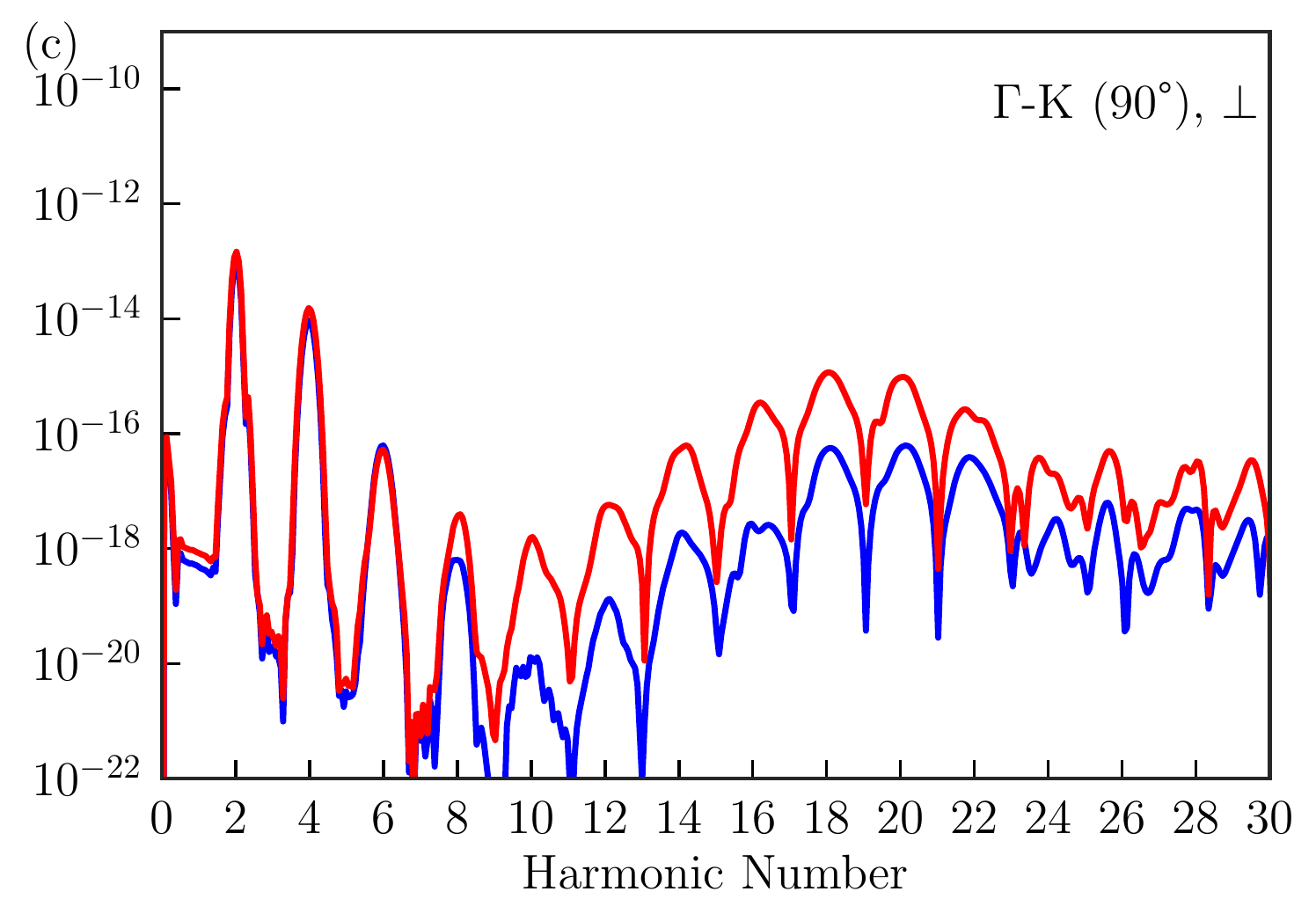}
\par\end{centering}
\raggedright{}\caption{\label{fig:5}High harmonic generation spectrum computed for a single
monolayer of hBN for the effective model (blue lines) and for the
\emph{ab-initio} model (red lines). (a,b,c) Harmonic spectrum for
a laser pulse in the $\Gamma-\mathrm{M}$ ($\Gamma-\mathrm{K}$,$\Gamma-\mathrm{K}$)
direction along its parallel (parallel, perpendicular) direction,
respectively.}
\end{figure*}

In Fig. \ref{fig:5}, it is shown the spectrum for different pulse
directions and different directions of emission. Symmetry imposes
that for the $\Gamma-\mathrm{M}$ direction of the laser pulse the
perpendicular current is strictly zero and for that reason is not
shown. On the other hand, for the parallel emission, both even and
odd harmonics are allowed due to the breaking of the inversion symmetry.
When the laser pulse is aligned along the $\Gamma-\mathrm{K}$ direction,
symmetry restricts harmonic emission to odd (even) harmonics for the
parallel (perpendicular) emission \citep{you2017anisotropic}. It
is clear that first order harmonics, that are generated mostly by
intraband current \citep{Vampa2014}, have an exponential decay until
harmonic 13, that corresponds approximately to the bandgap energy,
$5.62$ eV. It can be observed that the agreement between both models
is extremely good given the simplicity of the effective model. For
intraband harmonics the agreement between both approaches is almost
quantitatively and for harmonics higher than the bandgap energy, usually
coming from interband polarizations, the results agree on the overall
shape having a discrepancy in its absolute value.

The quantitative discrepancy in the region of interband harmonics
arises due to the approximation done in the construction of the dipole
operator in the effective model. The assumption made in Eq. (\ref{eq:digonal_r})
for the effective model strongly reduces the possibility of recombination
between distant Wannier orbitals and this can be observed by looking
at the suppression of high harmonics when comparing against the \emph{ab-initio}
model. Furthermore, the low order harmonics have a much better agreement
since it is not expected that recombination elements have a strong
influence on them, just like Brunel harmonics in atomic HHG.

\section*{Conclusion\label{sec:Conclusion}}

The Wannier gauge is presented in this work as a way to avoid the
problem of random phases of the dipole couplings along the FBZ. The
dynamical equations for the RDM are derived in the Wannier gauge.
The presented formalism can provide a new framework in which DFT calculations
and simple tight-binding Hamiltonians can be used for the description
of the HHG process. The harmonic spectrum for a single monolayer of
hBN is calculated and the results validate the use of the simplest
tight-binding Hamiltonian commonly used \citep{neto2009electronic}
to study HHG when compared to \emph{ab-initio} DFT calculations. Our
work opens the window for the calculation of nonlinear optical response
in crystals by using parameters directly extracted from \emph{ab-initio}
calculations using the MLWF procedure, that up to now was a cumbersome
task due to the problem of the random phases in the dipole couplings.
\begin{acknowledgments}
R.E.F.S. acknowledge fruitful discussions with Bruno Amorim and Álvaro
Jiménez-Galán. M.I. and R.E.F.S. acknowledge support from EPSRC/DSTL
MURI grant EP/N018680/1. R.E.F.S. acknowledges support from the European
Research Council Starting Grant (ERC-2016-STG714870).
\end{acknowledgments}

\end{document}